\documentstyle[11pt,amsfonts]{article}

\newcommand{\be}{\begin{equation}}
\newcommand{\ee}{\end{equation}}
\newcommand{\bea}{\begin{array}}
\newcommand{\ea}{\end{array}}
\newcommand{\beqa}{\begin{eqnarray}}
\newcommand{\eeqa}{\end{eqnarray}}
\newcommand{\bean}{\begin{eqnarray*}}
\newcommand{\eean}{\end{eqnarray*}}

\def\up#1{\leavevmode \raise.16ex\hbox{#1}}

\newcommand{\gapproxeq}{\lower
 .7ex\hbox{$\;\stackrel{\textstyle >}{\sim}\;$}}
\newcommand{\lapproxeq}{\lower .7ex\hbox{$\;\stackrel
{\textstyle <}{\sim}\;$}}

\newcounter{appendice}

\def\thebibliography#1{{\bf REFERENCES\markboth
 {REFERENCES}{REFERENCES}}\list
 {[\arabic{enumi}]}{\settowidth\labelwidth{[#1]}\leftmargin\labelwidth
 \advance\leftmargin\labelsep
 \usecounter{enumi}}
 \def\newblock{\hskip .11em plus .33em minus -.07em}
 \sloppy
 \sfcode`\.=1000\relax}

\begin{document}

\centerline{ \LARGE Charge Enhancement of Noncommutative Gravity }

\vskip 2cm

\centerline{   Mansour Haghighat$^{a,b}$\footnote{mansour@cc.iut.ac.ir} and Allen Stern$^{a}$\footnote{astern@bama.ua.edu}   }

\vskip 1cm
\begin{center}
  {\it a) Department of Physics, University of Alabama,\\ Tuscaloosa,
Alabama 35487, USA\\}
{\it b) Department of Physics, Isfahan University of Technology,\\ Isfahan 84156-83111, Iran\\}
\end{center}
\vskip 2cm

\vspace*{5mm}

\normalsize
\centerline{\bf ABSTRACT}

Noncommutative corrections to the  metric tensor  can be significantly enhanced by the presence of electromagnetic fields.  Neutron stars, with their large magnetic fields, are possible candidates to search for such effects. We use precision measurements of the gravitational redshift from pulsars to put bounds on the parameters of noncommutative gravity.

\newpage

\section{Introduction}

Noncommutative gravity, which corresponds to the noncommutative generalization of the Einstein field equations,  offers the possibility of modeling  effects from quantum gravity.
 Noncommutative dynamics is  commonly  obtained by replacing  pointwise products in the commutative equations of motion by noncommutative  star products.  The procedure has been applied to obtain the field equations for noncommutative gravity.  The procedure of replacing  pointwise products by noncommutative  star products is highly ambiguous, and consequently different formulations can be  given.  (For the case of gravity, see for example \cite{Chamseddine:2004si}, \cite{Aschieri:2005yw}.)   Under certain conditions, the  solutions to any commutative field equations are stable under noncommutative deformations obtained using the star product, and this  result is independent of the formulation.  This allows for exact solutions to the noncommutative  Einstein equations.\cite{Schupp:2009pt},\cite{Ohl:2009pv},\cite{Aschieri:2009qh},\cite{Dolan:2006hv}  
To give a physical interpretation to the noncommutative solutions  one can map them back to the  commutative theory, where familiar quantities  like the metric tensor  can be defined.  The map, known as  the Seiberg-Witten map\cite{Seiberg:1999vs}, can be applied to any gauge theory. Then upon writing gravity as a gauge theory, which is done in the formulation of \cite{Chamseddine:2004si}, one  obtains corrections to the metric tensor  for the commutative solutions. Noncommutative corrections to pure gravity solutions are second order in the noncommutative parameters $\Theta^{\mu\nu}$ defining the deformation, and as a result offer little hope   for direct observation.

The situation is different, however, for exact solutions to the noncommutative   Einstein-Maxwell equations,  obtained as before by replacing pointwise products by star products. If one formulates it as a gauge theory, with  electromagnetism and gravity contained  in a single gauge group, as was actually necessary in the formalism in  \cite{Chamseddine:2004si},
 then the  electromagnetic and Lorentz spin connections get mixed by the  Seiberg-Witten  map.  This means that there are 
 corrections  to the  metric tensor which are due to the electromagnetic fields, and corrections    to the electromagnetic   potentials  which are due to gravity.   Here the leading corrections are {\it  first order} in   $\Theta^{\mu\nu}$.\cite{Stern:2009id} For the example of a charged black hole, these  corrections to the metric tensor are also linear in the charge $Q$.  To write down their form,  it is convenient to introduce the dimensionless quantities   
 \be q= \frac Q{\sqrt{G}M} \label{smallq}\;,\qquad \quad \epsilon^{\mu\nu}=\frac{\Theta^{\mu\nu}}{G^{3/2}M}\;, \label{epmunu}\ee
where $M$ is the  mass and we assume natural units $c=\hbar=1$ throughout. The maximum (or extremal) value for $q$ is one.   Noncommutative corrections to the metric tensor go like $q\epsilon^{\mu\nu}$ and are thus significant when the product is of order one.    For the example where $M $ is a solar mass,  one has $$G^{3/2}M_{\odot}\approx\; {\rm GeV}{}^{-2}$$  This means that for  solar mass objects with close to the extremal charge $q\;{}^<_\sim\;1$, there would be significant noncommutative effects for  $\Theta^{\mu\nu}$ is of order $ \;$GeV${}^{-2}$. It makes it possible to obtain bounds on the parameters of noncommutative gravity from astronomical observations, which are comparable to bounds found on noncommutative parameters  in other domains, such atomic and high energy physics. (See, for example, \cite{bounds})  For this, we note that 
neutron stars have a nonvanishing charge, which is screened by its magnetosphere.\cite{Punsley},\cite{Michel}  Although $q$ may not be close to one for typical neutron stars, we can obtain an interesting bound on $\Theta^{\mu\nu}$ from precision measurements of the gravitational redshift.

For the application to neutron stars, a reasonable framework is the  Kerr-Newman solution to the Einstein-Maxwell equations, as it approximates the exterior  fields of the  star, and  includes some standard corrections to the gravitational redshift.  When $\Theta^{ij}=0$,  $i,j=1,2,3$, associated with the space components, it is also a solution of the noncommutative field equations. We compute the leading  noncommutative corrections to the  Kerr-Newman metric tensor due to  the remaining  components of $\Theta^{\mu\nu}$ in section two.  The results are applied to the case of millisecond pulsars and magnetars  in section three to obtain bounds on the noncommutativity parameters from  measurements of the gravitational redshift.

\section{ Kerr-Newman solution and its noncommutative corrections}

\setcounter{equation}{0}
The Kerr-Newman black hole is characterized by mass $M$, charge $Q$ and angular momentum per unit mass $a$.
For convenience, we shall write down the solution and its noncommutative corrections in terms of dimensionless  coordinates $\xi^\mu$, $\mu=0,1,2,3$, where
 $\xi^0=\tilde t =\frac t{GM}$ and  $\xi^1=\tilde r =\frac r{GM}$,  with $t$ being the time  and $r$ the radial variable, along with  the usual angular variables  $\xi^2=\theta$ and $\xi^3=\phi$,  $0\le\theta\le \pi$,  $0\le\phi\le 2\pi$.  From the  invariant  interval $d s^2=g_{\mu\nu}(\xi)d\xi^\mu d\xi^\nu$  and potential one form $ A=A_\mu(\xi) d\xi^\mu$,
one can  construct the dimensionless objects
\be d\tilde s^2\;=\;\frac{ds^2}{(GM)^2} \qquad\quad  \tilde A \;=\;
 \frac A{\sqrt{G}M} \;,\ee
 which for the Kerr-Newman solution can be written as
 \beqa d\tilde s^2&=& \left({-\tilde\Delta + \tilde a^2 \sin^2\theta}\right)\frac{d\tilde t^2}{\tilde \rho^2} + \frac{\tilde \rho^2}{\tilde{\Delta}}
 d\tilde r^2 +\tilde \rho^2 d\theta^2\cr & &\cr &+ &\frac{ \sin^2\theta}{\tilde \rho^2}\;\biggl\{ \Bigl({(\tilde r^2 +\tilde a^2)^2 -\tilde\Delta \tilde a^2 \sin^2\theta}\Bigr) d\phi -2 \tilde a (\tilde r^2 + \tilde a^2 -\tilde \Delta) d\tilde  t\biggr\}d\phi\label{knnvrntmtr}\\
& &\cr \tilde A & =&- \frac {q\tilde r}{\tilde \rho^2}\;(d\tilde t - \tilde a \sin^2\theta\; d\phi)\;,\label{knfstrngth} \eeqa
 where
\be \tilde \Delta=\tilde r^2 + \tilde a^2  +q^2 - 2\tilde r \qquad \quad\tilde \rho^2 = \tilde r^2 + \tilde a^2 \cos^2\theta\;,\label{dlta}\ee
and  $\tilde a$ is the rescaled  angular momentum density, $ \tilde a =\frac a{GM}$, and like $q$, its  maximum (or extremal) value  is one. 
The solution is valid 
 provided \be  q^2 + \tilde a^2\le 1 \label{bndMQa}\;\ee 
 
 Now let us introduce another set of coordinates $ x^\mu$ on space-time, which here have units of length. 
In passing to a noncommutative  theory, one standardly replaces  $ x^\mu$ by operators $\hat x^\mu$, satisfying some commutation relations
\be [\hat x^\mu ,\hat x^\nu] =\Theta^{\mu\nu} \label{fndmntlcm}\;,\ee $\Theta^{\mu\nu} $ having units of length-squared.  It is often convenient 
to realize the algebra on commutative space-time by introducing a star product $\star$, such that 
\be x^\mu\star x^\nu - x^\nu\star x^\mu=\Theta^{\mu\nu} \label{staralgebra}\ee 
We shall assume that there is a choice of  $x^\nu$, at least locally, whereby  $\Theta^{\mu\nu}$ is central in the algebra and independent of  $ x^\nu$.   Then $\star$ can be taken to be  the Groenewold-Moyal star product \cite{groe}. 
 
 Noncommutative 
field equations are standardly  obtained by replacing  pointwise products in the commutative in the field equations by star products. For certain choices of $\Theta^{\mu\nu}$, solutions of the commutative theory are also  solutions of the noncommutative theory.  
 This is the case for (\ref{knnvrntmtr}) and (\ref{knfstrngth}), or any
 any static solution, when we assume
  $\Theta^{ij}=0$ and we identify $x^0$ with the time $t$.  This follows because then
 the Groenewold-Moyal star product acting between time-independent functions reduces to the pointwise product of the functions, and a static solution of the commutative field equations,  is also a solution in the noncommutative field equations. So when $\Theta^{ij}=0$, (\ref{knnvrntmtr}) and (\ref{knfstrngth}) are solutions to {\it any} noncommutative Maxwell-Einstein equations obtained by replacing pointwise products by Groenewold-Moyal star products. 

As stated in the introduction,  the Seiberg-Witten map\cite{Seiberg:1999vs} can be applied to find the image of the solution in the commutative theory, and thereby obtain corrections to the  metric tensor and potentials of the commutative solution.  Starting with a gauge theory description which  contains both gravity and  electromagnetism in a single gauge group, the leading order
 corrections  $\delta_{\mbox{\tiny NC}}   g_{\mu\nu}$ to the  metric  are due to the $U(1)$ fields, and  the leading order corrections   $\delta_{\mbox{\tiny NC}}   A_{\mu}$ to the  potentials  are due to gravity.
Here we shall only be concerned with the former.  Applying the $GL(2,C)$ gauge theory of Chamseddine \cite{Chamseddine:2004si},  $\delta_{\mbox{\tiny NC}}   g_{\mu\nu}$  are first order in $\Theta^{\mu\nu}$ when electromagnetic fields are present in the solution.  The effect of the corrections is to translate the  space-time coordinates $x^\mu$ by $\Theta^{\mu\nu} A_\nu$.\cite{Stern:2009id}  That is,
\beqa  \delta_{\mbox{\tiny NC}}  g_{\mu\nu}& =&-\Theta^{\rho\sigma} A_\rho \frac{\partial g_{\mu\nu}}{\partial x^\sigma}+{\cal O}(\Theta ^2)\;\cr & &\cr & =&-\Theta^{ti} A_t \frac{\partial g_{\mu\nu}}{\partial x_i}+{\cal O}(\Theta ^2)\;,
\label{frrdrcrmtr} \eeqa 
using  $\Theta^{ij}=0$.

It remains to identify the spatial coordinates $x^i$ for the Kerr-Newman solution and  compute the noncommutative corrections (\ref{frrdrcrmtr}).  For the former, we shall assume that $x^i$ span $R^3$, and require that (\ref{staralgebra}) defines a self-adjoint algebra. This can be maintained for 
 \be \frac{ x^1}{GM}= \sqrt{\tilde r^2+\tilde  a^2} \sin\theta\cos\phi \qquad \frac{x^2} {GM}=\sqrt{\tilde r^2+ \tilde a^2} \sin\theta\sin\phi \qquad
 \frac{ x^3}{GM}=\tilde  r \cos\theta   \ee
 In computing the noncommutative corrections, we choose only $\Theta^{t3}$ nonzero, for simplicity,  which for the Kerr-Newman solution means that  the noncommutative direction is parallel to the rotation axis.  The leading order correction to invariant measure  (\ref{knnvrntmtr}) is of the form
\be  \delta_{\mbox{\tiny NC}} d\tilde s^2\;=\;\;\frac{{q\epsilon^{t3} } \tilde r \cos\theta}{\tilde \rho^4} \;\kappa_{\mu\nu} d\xi^\mu  d\xi^\nu  \;,\label{nccttlds2}
\ee
where $ \epsilon^{\mu\nu}$ is given in (\ref{epmunu}) and $\kappa_{\mu\nu}(\xi)$ are  given in the appendix.  $\kappa_{\mu\nu}(\xi)$ are well behaved at $\theta=\pi/2$, and so from  (\ref{nccttlds2}), noncommutative effects  vanish at the equator.  Then the geodesics orbits in the equatorial plane are unchanged at first order. [This situation will change if we allow for other components of $\Theta^{ti}$.]  
 The noncommutative correction  (\ref{nccttlds2}) can be approximated by
\be  \delta_{\mbox{\tiny NC}} d\tilde s^2\;\approx\;\;{2q\epsilon^{t3} } \cos\theta\biggl\{-\frac{d\tilde t^2}{\tilde r^3}-\frac{d\tilde r^2}{\tilde r(\tilde r -2)^2}+d\theta^2\biggr\}\;,\label{aprxnccrtn}
\ee  when
  $\frac{a^2}{R^2}<<1$ and   $\frac{Q^2}{MR}<<1$.
 These are reasonable assumptions for neutron stars, as is evident in the next section.
\section{Application to neutron stars} 
 \setcounter{equation}{0}
 
\subsection{Corrections to the gravitational redshift}

The gravitational redshift, along with  corrections,  can be read off  the $g_{tt}$ component of the metric tensor in (\ref{knnvrntmtr}):
\beqa g_{tt}+1 &=&  \frac{ 2\tilde r -q^2}{\tilde \rho^2}\cr & &\cr &\approx& \frac 2{\tilde r} - \frac {2\tilde a^2}{\tilde r^3} \cos^2\theta -\frac {q^2}{\tilde r^2}   \;,\label{rdplscrtn} \eeqa using $\frac{\tilde a^2}{\tilde r^2}<<1$. 
The first term on the right hand side is the standard gravitational redshift $ \delta _{\mbox{\tiny RS}}  g_{tt}$, while the remaining terms $\delta _{\mbox{\tiny J}}  g_{tt}$  and $\delta _{\mbox{\tiny Q}}  g_{tt}$ are corrections, respectively, associated with the angular momentum and charge 
\be 
\delta _{\mbox{\tiny RS}}  g_{tt} = \frac{2GM}R  \qquad  \delta _{\mbox{\tiny J}}  g_{tt}= - \frac{2GMa^2}{R^3} \cos^2\theta \qquad \delta _{\mbox{\tiny Q}}  g_{tt}= -\frac {GQ^2}{R^2}\;,\ee
 where $R$ is the radius of the source.  $\delta _{\mbox{\tiny J}}  g_{tt}$ is responsible for line broadening from the dependence on $\theta$.  There are additional corrections which are due to nonspherical deformations of the star and  the  Doppler effect, which produce both  frequency shifts and line broadening.\cite{kapoor} 
 From (\ref{aprxnccrtn}), the noncommutative correction to the gravitational red shift is
 \be  \delta _{\mbox{\tiny NC}}  g_{tt}\approx -\frac{2q\epsilon^{t3} \cos\theta}{\tilde r^3}= -\frac{2G MQ\Theta^{t3}\cos\theta}{R^3}\;,\label{dltancg00}
\ee
where we again assume    $\frac{a^2}{R^2}<<1$ and   $\frac{Q^2}{MR}<<1$. This would yield  an additional contribution to  line broadening. 

Below we examine  the corrections $\delta _{\mbox{\tiny J}}  g_{tt}$, $\delta _{\mbox{\tiny Q}}  g_{tt}$ and $\delta _{\mbox{\tiny NC}}  g_{tt}$ for neutron stars, assuming  $R\sim 10^6\;{\rm cm}$ and $M\sim M_{\odot}$. Neutron stars can be classified by their rotational frequencies $\Omega$  and magnetic field.  Millisecond pulsars are typically associated with magnetic fields of the order of   $10^{8}$ Gauss.   Precision measurements have been obtained for their gravitational red shift, from which we can get bounds on  $|\Theta^{t3}|$.
Magnetars have a lower frequencies and much larger magnetic fields, and can potentially yield better bounds.

\subsection{ Millisecond pulsars}

 We first consider rotational effects.  From   $\Omega\sim 10^3\; $sec${}^{-1}$, one has 
$a\sim R^2\Omega\sim 10^5\;{\rm cm}
$.  This is close to the extremal limit, $\tilde a\sim 1$, and the angular momentum correction to the redshift is of order $a^2/R^2\sim 1\%$. 

The effects due to the charge $Q$  for  a millisecond pulsar star are typically much smaller.  Here we first need an estimate of $Q$.
For this we assume  that the fields can be modeled by the Kerr-Newman solution for $r\ge R$ (although pulsars require that the magnetic poles are not be aligned with the rotational axis).   $Q$ is then determined by the magnetic field  from (\ref{knfstrngth}).
 Computing the surface magnetic field  directed along $x_3-$axis at the north pole $\theta=0$, one gets\cite{Punsley}  
\be B_3|_{\theta=0}= \frac{2QaR}{(R^2 +a^2)^2}\approx \frac{2Qa}{R^3}\label{bthree}\;,\ee
and we again assume  $\frac{a^2}{R^2}<<1$.
  The charge corresponding to $B_3|_{\theta=0}\sim 10^{8}$ Gauss$\;\sim\;10^{-12}\;$GeV${}^{2}$ is then
\be  Q\approx \frac{R^3 B_3|_{\theta=0}}{2a} \sim 10^{12}\;  {\rm C} \;,\label{QRB2a}\ee
or $\sim 10^{30}$ in natural units. This coincides with the value quoted for the Crab pulsar\cite{Michel}.   It gives a very tiny charge correction to the redshift, $\frac{Q^2}{MR}\sim 10^{-17}$, and   $q\sim 10^{-8}$ is far from the extremal limit.

From (\ref{dltancg00}), the line broadening  due to   noncommutativity is  
$\approx 4q\epsilon^{t3}/\tilde r^3$.  The gravitational redshift of millisecond pulsars has been measured with a precision of  the order of $.1\%$,\cite{Cottam:2002cu}
 and so one has \be
\frac {2q|\epsilon^{t3}|}{\tilde r^2}\; {}^<_\sim \;.001  \;\ee  We then find $|\epsilon^{t3}| \; {}^<_\sim \; 10^6$, and thus 
\be |\Theta^{t3}|\; {}^<_\sim \; {\rm MeV}^{-2} \label{bndonthta}\ee  

\subsection{Possible application to magnetars}

Magnetars have magnetic fields of   $10^{14}\;$Gauss and $\Omega\sim \; $sec${}^{-1}$.
From the latter, one gets  $a\sim 10^2\;$cm.  So here the angular momentum correction to the redshift is small  $a^2/R^2\sim 10^{-8}$ and    $\tilde a\sim 10^{-3}$ is far from the extremal limit.  
Upon substituting $B_3|_{\theta=0}\sim 10^{14}$ Gauss$\;\sim\;10^{-6}\;$GeV${}^{2}$ in (\ref{QRB2a}), we get $
Q \sim 10^{19}\;$C, or $\;\sim\;10^{37}$ in natural units.   
 This yields a larger charge correction to the redshift, $\frac{Q^2}{MR}\sim 10^{-3}$, and   $q\sim .1$ is close to the extremal limit. The lower limit on $|\epsilon^{t3}|$ is now $10^3$ times the precision of a measurement, and so if the gravitational red shift can be measured to the previous accuracy of $.1\%$, the resulting  bound on  $|\Theta^{t3}|$ would go down to $ {\rm GeV}^{-2}$.

We lastly remark that for the determination of the charge for magnetars is not very clear.
The above calculations for $Q$ using  (\ref{bthree}) may be questionable because it ignores internal and surface effects of the star, as well as effects from the magnetosphere. An indication of this is that, from  (\ref{bthree}), the extremal case $q=1$ (with  $\tilde a\sim 10^{-3}$)  corresponds to a surface magnetic field of  $10^{15}\;$Gauss, and stars having larger magnetic fields are thought  possible.\cite{Duncan2} $10^{15}\;$Gauss is also orders of  magnitude less than the upper bound obtained from the virial theorem of magnetohydrostatics equilibrium.\cite{frme}  An alternative formula for the charge, which takes into account internal and surface contributions of the star, is\cite{Punsley}  
\be
Q\approx\frac1{3}{\Omega R^3B_0}\;,\label{qfntsz}
\ee Contrary to (\ref{QRB2a}), it is proportional to the rotation frequency, and so $Q$ is suppressed for smaller $\Omega$.  In deriving this result, the charge density  and magnetic field   $B_0$ are  assumed to be uniform inside   the neutron star, and here, general relativity is completely ignored!  It yields  $
Q \sim 10^{11}\;$C for 
$\Omega\sim \; $sec${}^{-1}$ and $B_0\sim 10^{14}$ Gauss, which is similar to what was found for millisecond pulsars.   Without a more complete formula which takes into account both finite size effects and general relativity, we  can then only conclude  that the true charge for magnetars lies between this value and $10^{19}\;$C.

\appendix

\section{Appendix}
$\kappa_{\mu\nu}$ in (\ref{nccttlds2}) are given by
\beqa  \kappa_{\tilde t \tilde t} &= &\frac{2}{\tilde \rho ^4} \biggl(\tilde a^2\left(2\tilde  r q^2-4 \tilde r^2+\tilde \rho ^2\right) -4 \tilde r^4+3 \tilde \rho ^2 \tilde r^2+q^2\tilde r \left(2 \tilde r^2-\tilde \rho
   ^2\right)\biggr)
\cr & &\cr
   \kappa_{\tilde r \tilde r} &= &\frac{2    }{\tilde \Delta } \biggl(\tilde r \left(\tilde r^2+\tilde a^2+\tilde a^2\sin ^2\theta \right)-\frac{\tilde \rho ^2(\tilde r-1) \left(\tilde r^2+\tilde a^2\right)
   }{\tilde \Delta}\biggr)
   \cr & &\cr 
   \kappa_{\theta\theta} &= & 2 \tilde  r   \left(2 \tilde a^2+2 \tilde r^2-\tilde \rho ^2\right) 
  \cr & &\cr 
   \kappa_{\phi\phi} &= & \frac{2 \tilde a^2   \sin ^4\theta }{\tilde \rho ^4}  \biggl(\tilde a^2\left(2 \tilde r q^2-4 \tilde r^2+\tilde \rho ^2\right) +\tilde r \left(q^2 \left(2\tilde  r^2+\tilde \rho ^2\right)-\tilde r \left(4
\tilde    r^2+\tilde \rho ^2\right)\right)\biggr) 
       \cr & &
  \cr  \kappa_{\tilde t\phi} &= & 
  -\frac{2\tilde  a   \sin ^2\theta }{\tilde \rho
   ^4} \left(\tilde a^2+\tilde r^2\right)   \left(2\tilde  r q^2-4 \tilde r^2+\tilde \rho ^2\right) \label{kappamunu}
   \eeqa

\end{document}